%% file: ms.tex
\def\LUM{\:{\rm ergs\:s^{-1}}}
\def\FLUX{\:{\rm ergs\:cm^{-2}\:s^{-1}}}

\def\VEL{\:{\rm km\:s^{-1}}}
\def\fuse{\:{FUSE}}

\documentclass[12pt,preprint]{aastex}

\begin{document}

% Additional private definitions that appear to work only inside document

\newcommand{\MSOL}{\mbox{$\:M_{\sun}$}}

% End of defining things

\title{ {\it Far Ultraviolet Spectroscopic Explorer} 
Survey of Magellanic Cloud Supernova Remnants\altaffilmark{1}}
\author{William P. Blair\altaffilmark{2}, Parviz Ghavamian\altaffilmark{2}, 
Ravi Sankrit\altaffilmark{2},
\& Charles W. Danforth\altaffilmark{3}
}

\altaffiltext{1}{Based on observations made with the NASA-CNES-CSA 
Far Ultraviolet Spectroscopic Explorer. $\fuse$ is operated for NASA 
by the Johns Hopkins University under NASA contract NAS5-32985.}

\altaffiltext{2}{Department of Physics and Astronomy, The Johns Hopkins
University, 3400 N. Charles Street,  Baltimore, MD 21218;
wpb@pha.jhu.edu, parviz@pha.jhu.edu, ravi@pha.jhu.edu}

\altaffiltext{3}{Center for Astrophysics and Space Astronomy,
University of Colorado, Boulder, CO 80309; danforth@casa.colorado.edu}

\begin{abstract}

We report the progress to date from an ongoing
unbiased ultraviolet survey of supernova remnants in the
Magellanic Clouds using the Far Ultraviolet Spectroscopic Explorer 
($\fuse$) satellite.  Earlier work with $\fuse$ and other instruments
has indicated that
optical and/or X-ray characteristics of supernova remnants are not 
always good predictors of their brightness in the ultraviolet.
This survey is obtaining spectra of a random large sample of Magellanic 
Cloud supernova remnants with a broad range of radio, optical, and X-ray 
properties.  We proposed 39 objects in the Large Magellanic Cloud and
11 objects from the Small Magellanic Cloud, with a standard request of
10 ks per object using the $\fuse$ 30\arcsec\ square aperture.  
To date, 39 objects have been observed in the survey (38 in the LMC 
and one in the SMC) and 
15 have been detected, a detection rate of nearly 40\%.
Our survey has nearly tripled the number of UV-detected
SNRs in the Magellanic Clouds (from 8 to 22).
Because of the diffuse source sensitivity of $\fuse$, upper limits
on non-detected objects are quite sensitive in many cases, dependent
upon night observing fraction and whether stellar light contamination
plays a role for a given object.  Estimated total luminosities in
O~VI, based simply on scaling the flux at the observed positions to 
an entire object, span a broad range from considerably brighter to many 
times fainter than the inferred soft X-ray luminosities, indicating
that O~VI can be an important and largely unrecognized
coolant in certain objects. 
We compare the optical and X-ray properties of the detected and 
non-detected objects but do not find a simple indicator for 
ultraviolet detectability. 
Non-detections may be due to clumpiness of the emission, high foreground
extinction, slow shocks whose emission gets attenuated by the Magellanic 
interstellar medium, or a combination of these effects.
The characteristics of individual detected 
supernova remnants are summarized in an Appendix.

\end{abstract}


\section{Introduction}

Supernova remnants (SNRs) as a class are important constituents of
the interstellar medium (ISM).  They are responsible for much of 
the enrichment of the ISM of a galaxy, and they are a significant 
energy source for stirring and mixing the ISM.  They may also be important
in some regions for triggering star formation in molecular clouds.
SNRs typically emit over nearly the entire electromagnetic spectrum,
and multiwavelength observations have become an important source of
physical information about the ejecta (in young objects), and the
circumstellar medium and ISM into which the shock wave propagates.

Ultraviolet spectroscopy of SNRs has proven to be an important tool for
providing physical information and insight into this important class
of objects (Blair 2001; Raymond 2001; and references therein). 
Emission lines in the UV tend to come from material at
temperatures that are intermediate between the X-ray emitting
gas and the optical filaments.  
Information on shock velocities, preshock densities, and abundances
of important elements such as carbon that have no bright optical lines
can all be derived from UV observations of shocks.  $\fuse$ observations
have the added bonus of high spectral resolution, which provides
kinematic and line profile information.  Thus the importance of overlying 
absorption and in some cases the shock geometry can be assessed directly 
(Blair, Sankrit, \& Tulin 2002; Sankrit, Blair, \& Raymond 2003).

Primarily because of interstellar extinction, however, there are
remarkably few objects available for study in the UV. In
our Galaxy, the Cygnus Loop and Vela SNRs are the objects that have
been studied extensively; both are nearby and have low foreground
extinction.  Notable but difficult observations of the Crab Nebula
(Blair et al. 1992), SN 1006 (Raymond, Blair, \& Long 1995; Korreck et al.
2004) and Puppis A (Blair et al. 1995) have also been performed.

The foreground galactic extinction toward the Magellanic Clouds is
relatively low, but the number of SNRs observed in the FUV is
still very modest (see Table 1).
The young `core collapse' SNRs N132D (LMC) and 1E 0102-7219 (SMC) 
are very bright in soft X-rays and have been
studied in the UV as well (Morse et al. 1996; Blair et al. 2000a). 
The X-ray and optically-bright SNR N49 has received considerable 
attention (Vancura et al. 1992; Blair et al. 2000b; Sankrit, Blair, 
\& Raymond 2004).  However, the next 
brightest optical SNR in the LMC, N63A, has
not been observed in the ultraviolet since the early days of the 
International Ultraviolet Explorer (IUE) satellite (Benvenuti, Dopita,
\& D'Odorico 1980) even though it should be 
readily detectable by modern instrumentation.  
Three out of four of the Balmer-dominated SNRs identified by Tuohy et al. 
(1982) have only recently been detected with $\fuse$ (Ghavamian et al. 2006).
Not only is this a small sample, but the objects observed have
tended to have specific characteristics that attracted attention
to them, and with very bright optical and/or X-ray
emission in particular.  They are not `typical' SNRs, and so the 
picture we derive for the global impact of SNRs
on their host galaxy is seriously biased.

Positions for FUV spectral observations have
often been selected based on the optical appearances or characteristics 
of the region or object being observed, including not only the surface
brightness but also particular line ratios or signatures (e.g., Raymond 
et al. 1988; Blair et al. 2002; and references therein).
More recently, a few FUV observation positions have been selected 
based on the appearance in soft X-rays rather than optical (Raymond et 
al. 1997; Sankrit et al. 2001). With very few FUV images available for 
guidance, observers have had little else to go by.
 
However, there have been a growing number of indications that
selection criteria for FUV observations of SNRs have been too narrow.  
FUV spectro-imaging of the Cygnus Loop and Vela SNRs from the Voyager
Ultraviolet Spectrometers (Blair et al. 1991, 1995) and of Vela recently
by the SPEAR experiment (Nishikida et al. 2006) show patchy, variable
spatial distributions in key FUV emission lines, albeit at modest
spatial resolutions. Higher spatial resolution FUV images of selected
regions in the Cygnus Loop are available from the Ultraviolet 
Imaging Telescope (UIT). Danforth et al. (2000) show comparisons of 
UIT images to both X-ray and optical images, indicating all wavelengths 
are patchy and show relative spatial variations.
Even though the UIT bandpass did not isolate individual emission lines,
spatial variations of the FUV emission are clearly present.
Some of the spectroscopic positions observed
in the Cygnus Loop and Vela SNRs also correspond to very faint optical 
filaments and yet the UV emission is bright (Sankrit et al. 2001; 
Blair et al. 2002; Sankrit \& Blair 2002; Raymond et al.  2003).  

A serendipitous $\fuse$ observation of a SNR in the SMC H~II region
NGC~346 (N66) has provided another important example of potential
observational bias (Danforth et al. 2003).
With the $\fuse$ MDRS aperture placed on a star in the NGC~346 star cluster,
strong O~VI and C~III $\lambda$977 emission lines were detected in 
the LWRS aperture a couple of arcminutes away.  Reconstructing the 
LWRS aperture location showed that it was projected onto the limb of a 
known radio SNR 0057-7226 (Ye, Turtle, \& Kennicutt 1991). The SNR had been
observed previously in FUV absorption with $\fuse$, using the Luminous Blue 
Variable HD~5980 as a background source (Koenigsberger et al. 2001; Hoopes
et al. 2001), but no FUV emission had been detected.  Indeed,
{\it no optical shell or emission from the SNR is evident in images
of the region}, and only extremely faint optical
emission from the SNR can be detected in H$\alpha$ echelle data
(Chu \& Kennicutt 1988; Danforth et al. 2003).
The SNR is an X-ray source (Wang \& Wu 1992; Yokogawa et al. 2000;
Naz\'{e} et al. 2002), but with a luminosity of only
$\rm L_{x}(0.3 - 10 ~ keV) \simeq 1.4 \times 10^{35} ~ \LUM$ 
(Naz\'{e} et al. 2002) is certainly not in the class that would make it
a priority FUV target.  It is unlikely that an FUV observation 
of this SNR would have ever been attempted had the serendipitous 
spectrum not been obtained, and yet this is a bright FUV emission source.  

To understand the systematics of ultraviolet emission from a wider
sample of SNRs, and in particular to investigate the properties
of more typical SNRs, we are conducting an FUV survey of Magellanic 
Cloud SNRs that is not biased by optical, radio,\footnotemark \,
\footnotetext{In general, radio observations of Magellanic Cloud
SNRs have had insufficient resolution for detailed comparisons
with optical or X-ray imagery, but this is changing. See Dickel
et al. (2005).}
or X-ray characteristics 
of the selected objects.  While the survey is ongoing, we have 
a significant subset of the proposed observations in hand that 
allows us to address
some of the  primary questions about FUV detectability.  Section 2 
addresses the target selection criteria and describes the survey 
strategy and observational details. Section 3 describes a general
analysis of the detected objects and compares to optical and X-ray 
data.  Details about the individual objects detected are described
in the Appendix.
  
\section{Target Selection and Observations}

The $\fuse$ Survey program category allows one to propose a set of
targets for potential observation, with no guarantees that any 
particular target will be observed.  Targets are selected and inserted 
into observation timelines in ways that provide scheduling
flexibility to the observatory, while accomplishing the 
scientific objectives of the proposed program.  We thus compiled a 
list of potential targets from the optical SNR catalogs of 
Mathewson et al. (1983; 1984, 1985) and from the X-ray SNR catalog 
of Williams et al. (1999a).  We supplemented
this list with a number of individual objects reported in the literature
(Smith et al. 1994; Chu et al. 1995; 1997; 2000; Dickel et al. 2001;
Smith et al. 2004), and a handful of objects where optical longslit echelle
data from CTIO (Danforth 2003) indicated
high velocity emission in H$\alpha$ but for which
no previous SNR identification had been suggested.  Such objects may
represent SNRs `buried' in H~II regions that have not heretofore called
attention to themselves, similar to the SMC SNR 0057-7226 mentioned earlier.  
We culled this list by removing objects previously
observed with $\fuse$ (see Table 1) and a small subset for which 
optical extinction
measurements or X-ray derived neutral hydrogen column densities
indicated severe attenuation of any potential FUV emission.
Finally, a selection of six SNRs newly identified in the Magellanic Cloud
Emission Line Survey (MCELS) team (Smith et al. 2004) were added in 
the second year of the survey.
Our final list contained 39 objects in the LMC and 11 objects in the SMC.
The observations reported below took place over two observing cycles, 
using $\fuse$ program identifiers D904 (44 LMC and SMC remnants)
and E900 (six new MCELS objects in the LMC).

The $\fuse$ instrument covers the wavelength range 905 -- 1187 \AA, with 
a nominal point source resolution R = $\lambda/\Delta\lambda ~ \ge$ 20,000.
For diffuse sources of interest to this program, the spectral resolution
is driven instead by the spectrograph aperture size and filling factor. 
We are primarily interested in the LWRS (30\arcsec\ square) apertures
for this paper.  A filled LWRS aperture produces a roughly square-topped
instrumental profile of width 106 $\VEL$ near 1035 \AA.
$\fuse$ contains four optical channels, each with its own focal plane 
and spectrograph aperture plate.\footnotemark 

\footnotetext{Note: channels are 
referred to as LiF1, LiF2, SiC1 and 
SiC2, where LiF and SiC refer to the optical coatings on each channel
and the numbers refer to one of two microchannel plate detectors.  
Furthermore, each detector is sub-divided into
two segments, A and B, whose boundaries in wavelength space are offset
slightly so that full wavelength coverage is maintained.  See section 3 of 
Moos et al. (2000) for full details.}

Small, thermally-induced distortions in the $\fuse$ optical bench 
discovered after launch do not allow 
rigid co-alignment of the apertures from each channel over time. 
Typical misalignments correspond to a few to as much as 10\arcsec,
and can vary even during the course of a given orbit or integration.  
For reference, the LiF1 LWRS aperture positions are shown in this paper.
The LiF1 channel was used for guiding, and so its position on the 
sky was held fixed while the other channels drifted relative to it. 
In cases where the object is large in comparison to the $\fuse$ apertures,
all of the apertures should be sampling similar SNR emission. However, 
on the smaller objects,
channel misalignments can cause different sampling and impact the
assessments of relative O~VI and C~III line strengths.
Further details about the $\fuse$ instrument and on-orbit 
performance information are provided by Moos et al. (2000; 2002) and Sahnow 
et al. (2000).

The $\fuse$ survey strategy involved a standard 10 ks request on each object.
These faint emission line sources could all be observed in time-tagged mode.
The size of the LWRS apertures plus the sensitivity of the $\fuse$ detectors 
combine to provide exceptional diffuse source sensitivity compared with 
other FUV spectrographs that have been flown. For instance, 
a nominal 10 ks observation would detect emission in 
O~VI $\lambda$1032 that is 100 times fainter than the bright LMC SNR N49
(Blair et al. 2000; Sankrit et al. 2004), with limiting flux levels near
$\rm 5 ~\times ~ 10^{-15} ~ ergs~cm^{-2}~s^{-1}$.
Actual observation times ranged above and below the nominal 
10 ks request, at the convenience of the $\fuse$ schedulers. (See discussion
below and Table 2.)

%In each case, the LiF1A LWRS aperture was placed at the projected center 
%of each SNR, as judged from the published catalog data.  
%No attempt was made to tailor the aperture
%positions based on observed morphology or structure.
The intent of the survey was to place the LiF1A LWRS aperture at or near
the projected center of each SNR.  This strategy had two primary
motivations.  First of all, to do anything other than this would
apply a bias to our survey that was not desirable.  We did not want 
to point at features selected from optical or X-ray images 
of the objects.  Secondly, it may be advantageous to point
through the centers, where typical Doppler shifting of $\pm$30 -- 150 
$\VEL$ would move any SNR emissions out from under potential overlying 
absorptions by the host galaxy.  This is an important factor for strong
resonance lines such as C~III $\lambda$977 and O~VI, the strongest 
lines seen in most SNR spectra.  

In practice, since published catalog coordinates were used for many of 
the objects, the $\fuse$ apertures did not always lie on the exact central 
positions. In objects with spotty optical emission, for instance, some of
the Mathewson et al. (1983, 1984, 1985) coordinates are off center
with respect to newer X-ray data, which tends to show the full extent
of more objects.  Also, in retrospect, some of the MCELS-supplied coordinates
did not correspond to the object centers.  In all of the SNR figures
that follow, we show the actual LiF1A LWRS aperture location observed
(i.e., coordinates shown in Table 2), so any miscenterings will be
obvious.  Since most of the objects in our survey were from 1\arcmin\ 
to 5\arcmin\ in extent, the LWRS apertures were usually filled with potential 
emission, even if modest channel misalignments or miscenterings were present.  
A number of the
spectra are impacted to some degree by stellar emission, which affects our
ability to sense the presence of potential faint emission features.
Strong emission, if present, is still visible even if modest stellar
contamination is present.

To date, 39 objects out of the 50 potential targets have been
observed, all but one (SNR 0104-723) in the LMC.  
The distribution of these objects across the face of the LMC in shown
in Figure 1, where detections and non-detections are indicated with
separate symbols.  Table 2 summarizes the targets observed and the 
results in terms of detections and non-detections.  The details
of the spectral analysis and search for SNR emission lines are described 
in the next section.

%\footnotetext{IRAF is distributed by the National Optical Astronomy
%Observatories, which is operated by the Association of Universities for
%Research in Astronomy, Inc. (AURA) under cooperative agreement with the
%National Science Foundation.}

\section{Analysis and Discussion}

Each $\fuse$ data set has been reprocessed with the most current version
of the CalFUSE pipeline publicly available at the time of our
analysis, CalFUSE 3.0.7.  The output 
from this version of CalFUSE makes it trivial to inspect both the total 
data sets and the orbital-night only time periods. The total data sets were
inspected first.  However, scattered light and backscattered solar
line emission can cause confusion in the total data sets, especially for
the SiC channels (which include the C~III $\lambda$977 line).  The
redshift of each Magellanic Cloud is enough to separate the SNR C~III 
emission (when present) from the back-scattered solar emission, but
potential faint emission can still be lost in the higher background. 
For those objects with detected C~III and/or O~VI line emission,
we have chosen to extract, plot, and measure the night-only fractions 
to avoid contamination problems when possible. (Two detected objects
had no orbital night integration, however.  See Table 2.)

The search for emission centered on the brightest transitions
normally seen in the FUV -- 
O~VI $\lambda\lambda$1031.9138,1037.6154\footnotemark \,
and C~III $\lambda$977.020.
\footnotetext{For convenience, we will sometimes refer to these lines 
in the text with abbreviated wavelength notation, viz. 
$\lambda\lambda$1032,1038.}
The O~VI $\lambda$1032 line in the LiF1 channel was the primary 
feature searched for 
because the LiF1 channel has the highest effective area and the LWRS
aperture position was known accurately.  For most detected objects,
the LiF2B channel also showed O~VI emission, but at lower signal levels 
(due to the smaller effective area of this channel).\footnotemark \,
\footnotetext{An exception is SNR 0509-687 (N103B), for which a strong 
O~VI signal was seen in LiF1A but no signal was seen in LiF2B.  Apparently 
the channel misalignment was significant enough on this small diameter SNR to
cause the LiF2B channel to miss the SNR entirely. In a similar manner,
we cannot be certain whether the absence of C~III for this object's spectrum 
is intrinsic or due to channel misalignments.}
O~VI $\lambda$1038 was an
important secondary indicator, although it is nearly always impacted
severely by overlying absorption.  
Comparison with published grids of shock models (e.g., Hartigan,
Raymond, \& Hartmann 1987) indicate that strong C~III emission
occurs for shocks with velocities above about 100 $\VEL$ while O~VI 
emission `turns on' above $\sim$160 $\VEL$, when the shock becomes 
effective at ionizing $\rm O^{+4}$ to $\rm O^{+5}$.  
Hence, we also searched for C~III, which 
in principle could be seen even if no O~VI
was detectable.  The SiC channels both cover the C~III line region, 
but the SiC2A channel has higher effective area, so this channel was
searched for SNR emission.  Of course, channel misalignment
makes the exact location of the SiC2 LWRS aperture uncertain 
relative to the LiF1 LWRS aperture, as described in Section 2.

Out of the 39 objects observed to date, 15 detections have been obtained,
as indicated in Table 2, a detection rate of 38.5\%.
Summary figures showing images of each of these objects and $\fuse$
data sections are shown in Figures 2-16.  Optical H$\alpha$ and [S~II] 
CCD images in these figures were provided to us by the 
MCELS team (Smith et al. 1999), headed by R. Chris Smith of NOAO. 
The X-ray images have been obtained from the public HEASARC data archive for
Chandra, ROSAT, XMM-Newton, and Einstein.  Squares in the image panels 
show the nominal LiF1A LWRS aperture position and orientation for each 
observation.  Since the LWRS aperture is 30\arcsec\ in size, this
also provides the spatial scale for each image.  $\fuse$ data segments 
near the O~VI and C~III regions are shown in separate panels of each
Figure.  Because we are searching for faint emission lines, we found it
beneficial to bin most data sets over 8 CalFUSE 3.0.7 pixels ($\sim$30 
$\VEL$) for display in the Figures.  A brief discussion of each 
detected SNR is presented in the Appendix.

As can be seen in Figures 2 - 16, the line profiles exhibit a wide 
range of shapes.  Some are broad (several hundred $\VEL$), some 
narrow (only slightly above the filled aperture resolution), and some 
are multi-peaked. 
The O~VI $\lambda$1032 profile is most instructive, as it is often the
strongest line seen and has relatively good signal-to-noise ratio.
These figures each contain  horizontal bars with tick marks 
that indicate the positions of {\it potential} ISM absorption features,
primarily $\rm H_2$ and O~I lines. The bar marked ``MW" (for Milky Way)
shows the Galactic rest frame, and the other bar is shifted to the rest
frame of the appropriate Magellanic Cloud (+275 $\VEL$ for LMC; +165 
$\VEL$ for SMC).  The longest tick marks 
indicate the expected positions of C~III and O~VI lines.
Of course, the same transitions that we expect from the SNRs in 
emission are strong resonance lines that can also be in absorption 
from the intervening ISM at both galactic and MC velocities.  

Many objects detected in our survey show both O~VI and C~III emission
at detectable levels.  Interestingly, some objects show O~VI but no 
C~III (e.g., SNR 0509-687), but there is only one detection in 
C~III-only (SNR 0527-658), and this detection is marginal.  
The presence of O~VI 
emission in most of the detected objects indicates that each of them 
has some shocks with velocities in excess of 160 $\VEL$. 
In SNR 0527-658, at least at the observed location, such shocks are
apparently absent.  Additional observations at other locations within 
SNR 0527-658 would be needed to understand whether the absence of O~VI 
is intrinsic to the object or due to the clumpiness of the O~VI 
spatial distribution.

We have measured the observed line widths, approximate centroid velocities,
and integrated total line 
fluxes for all SNRs with detections, and list these values in Table 3.
These measurements have been made with IDL tools that permit user
interaction in the measurements. Each spectrum is displayed and a cursor
is used to set limits around each line for integrating the flux above
the background for each feature. Since most of the line profiles are
distinctly non-Gaussian, the fluxes derived in this way are simply
integrated flux above background, not the equivalent flux for a Gaussian fit.
These measurements are only lower limits to the
actual fluxes since unknown but potentially significant amounts of 
emission are apparently absorbed by foreground ISM. (See discussion below.)
The full-width-zero-intensity (FWZI) values listed in Table 3 indicate the full
range over which the integration was made.  The values listed in Table 3
are all significantly above the value expected for a filled LWRS aperture,
indicating that intrinsic line broadening is detected in all cases.
The central velocities listed are not the centroid of measured flux, 
but rather the center of the FWZI velocity range used.
%Thus far the faintest detections in our survey are at flux
%levels F(1032)\,$\approx$\,10$^{-14}$ ergs cm$^{-2}$ s$^{-1}$ and 
%F(977)\,$\approx$\,1.4$\times$10$^{-14}$ ergs cm$^{-2}$ s$^{-1}$.  

Two things are immediately obvious from inspection of Table 3.  For all objects
with both O~VI and C~III lines, the central velocities for C~III are more
redshifted than for O~VI, and the FWZI line widths are significantly higher
for O~VI than for C~III.  Both of these facts are consistent with significant 
impacts from overlying absorption.  Inspection of any of Figures 3 - 59 
from the Danforth et al. (2002) $\fuse$ ISM atlas of sight lines in 
the Magellanic Clouds shows that the C~III ISM absorption at LMC velocities
is typically very saturated and broader than the ISM absorption from O~VI. 
Hence, the portion of the C~III SNR emission that is visible from under 
this ISM absorption is narrower and appears more redshifted.  While O~VI 
ISM absorption is also usually present, it is not usually saturated and 
varies in strength between sight lines.  In the SNRs, often a broad base 
of emission is seen in O~VI 
$\lambda$1032 (for example, SNR 0454-665, Figure 5), and thus a larger 
FWZI is measured and the central velocity is less shifted than for C~III. 

Overlying absorption is also obviously present in many cases from inspection
of the line profiles themselves.  There are some cases where an apparently 
broad O~VI emission line has been partially but not totally absorbed on the
blue side by LMC and/or galactic ISM (e.g., SNR 0506-680, Figure 6).
In general, the line widths for C~III are equal to or narrower than 
widths for O~VI, consistent with a saturated C~III ISM absorption (even 
stronger than in O~VI) impacting the observed profile.  
The line profiles often give the impression of being cut off on 
the blue side, as if they are peeking out from under the overlying 
LMC absorption.  We see only one example
where substantial emission is seen blueward of the rest LMC (or SMC)
wavelength, that being SNR 0536-706 (Figure 15), and in this case the
blue wing of the line appears to be truncated by Milky Way O~VI absorption.  
Since we are looking through the 
middle of each SNR, we are preferentially seeing redshifted emission 
from the back side of each SNR, while the front-side shell emission is 
typically absorbed or partially absorbed by MC and Milky Way ISM absorptions.

We now address the one object with detected C~III and no O~VI emission,
SNR 0527-658 (Figure 9).
It is conceivable that O~VI emission is present in this object
but just not at the location of the LWRS aperture, or that this
object is dominated by shocks slower than $\sim 160 \VEL$. We have
no independent way of checking which of these possibilities may apply
in this case.  If the former suggestion is correct, it would imply
a rather clumpy distribution of O~VI, at least in some objects.
If the latter situation is true, it may provide a reason why
few (only one to date)  C~III-only objects are observed.
Assuming we are systematically seeing the redshifted sides of
the shells, objects with slower shocks (and hence lower bulk motions)
will tend to be buried (or mostly buried) underneath the strongly
saturated MC C~III ISM absorption line.
We note that the observed C~III profile in the spectrum of SNR 0527-658
appears to be cut off sharply on the blue side. Apparently in this case,
the relative motions are such that the C~III line is partially seen
in spite of the overlying absorption. 

In general, however, a combination of effects tend to complicate
the situation with C~III.  The uncertainty in position of the SiC channels
with respect to LiF1, the lower effective area, and the propensity for
significant spectral comtamination by solar-backscattered C~III all
work against C~III detection at various levels. (See more detailed 
discussion below for Figure 17.)

Non-detections
come in two basic varieties, those for which no significant evidence
for O~VI or C~III emission lines is seen, and those observations for which
significant stellar light contaminates the $\fuse$ data and makes 
detection of faint emission lines difficult to impossible. Figure 17 
shows the C~III and
O~VI regions of $\fuse$ data for SNR 0507-685, one of the non-detections
without significant stellar contamination.
This example shows both total and night-only data segments
for comparison.  No O~VI lines are detected, and while some spectral
structure appears at nearly the correct wavelength for LMC C~III emission,
the large change between the total and night-only data sets implies
the observed feature is not intrinsic SNR emission. In the case of the
SiC2A channel, this is an instrumental feature caused by the miscentering
of low pulse-height photons from the strong airglow feature to the left.
Still, the presence of such structure in the spectrum limits the 
ability to set a significant upper limit at C~III unless a significant
amount of orbital-night exposure time is available. 

Figure 18 shows the same spectral regions for SNR 0519-697 (N120), one 
of the objects with stellar contamination. 
This example shows that a peak near the expected O~VI position is present
at a low level, but its interpretation is uncertain. Many hot stars
contain P-Cygni wind features of varying strengths and shapes
that could produce some O~VI emission.  On the other hand, some
SNR emission could be present at an undetermined level and be masked
by the variable stellar continuum.  In cases such as this (indicated 
in Table 2), our ability to place limits on O~VI are made inherently 
more difficult.

Thus, the quality of the upper limits varies based on the achieved 
integration times for each object and whether stellar contamination is
present.  Also, the unknown velocity widths for the undetected objects
makes placing a specific upper limit rather arbitrary.  We choose to
adopt the method of Dixon, Sankrit, \& Otte (2006), who have determined
upper limits for a wide range of $\fuse$ data empirically, assuming a width 
appropriate for a filled LWRS aperture and the LiF1 channel, viz.

\begin{equation}
\rm U(1032) ~=~ \frac{273.1}{\sqrt{t_{exp}}}
\end{equation}

\noindent
where $\rm t_{exp}$ is the night exposure time in s, and U(1032) is a 
3$\sigma$ upper limit at $\lambda$1032 in units of 1000 Line 
Units\footnotemark \, (KLU),
\footnotetext{A Line Unit corresponds to 1 
photon $\rm cm^{-2} ~ s^{-1} ~ sr^{-1}$.}
where 1 KLU = $\rm 4.5 \times 10^{-19} ~ ergs ~cm^{-2} ~ s^{-1}~arcsec^{-2}$.
Thus, a night integration of 5 ks corresponds to U(1032) = 3.9 KLU or
a 3$\sigma$ LWRS upper limit of $\rm 1.56 \times 10^{-15} ~ ergs ~cm^{-2} 
~ s^{-1}$, assuming a filled aperture and an unresolved line width
(106 $\rm km ~ s^{-1}$).  This is more than two orders of magnitude below
the measured F(1032) in the bright LMC remnant N49 (Blair et al. 2000b), 
although the line width in that case was over 400 $\VEL$.

Few of the non-detections can be attributed to short integration times
(see Table 2), and the inferred upper limits on emission in O~VI, at 
least at the observed locations, are quite small. 
Of course in many
objects the region sampled by the $\fuse$ LWRS aperture is a tiny part of
the projected surface area.  A clumpy distribution of O~VI emission 
could be present, so the absence of detected O~VI does not necessarily
indicate a complete absence of O~VI emission from an object.  
For completeness, we indicate the $\fuse$ aperture positions on optical 
(MCELS) and X-ray images (when available) of all non-detections in 
Figures 19-24.  
Inspection of these Figures relative to the detections (Figures 2-16)
demonstrates that, judging from optical and X-ray morphology alone. 
one would be hard pressed 
to determine {\it a priori} which objects will be detectable 
in the FUV and which will not.  For instance, some large, old optical shell
remnants are seen (SNR 0450-709, Figure 2) and some are not (SNR 0453-672,
Figure 19, panel 3).  Perhaps more telling, several of the most solid 
detections with $\fuse$ are unimpressive optical and/or X-ray SNRs (e.g.,
SNR 0454.5-6713, Figure 4; SNR 0506-680, Figure 6) while some non-detections
are relatively strong optical and/or X-ray sources (e.g., SNR 0455-687,
Figure 20 top panel; SNR 0519-697, Figure 21 second panel).  There are
several SNRs with apparent central X-ray emission consistent with ejecta
in both the detected and non-detected samples, but it is not clear we are 
seeing FUV emission associated with the ejecta in any case, so this does
not appear to be a significant factor in FUV detectability.  

In Figures 25 and 26, we study some systematics of the sample. In Figure
25, we show histograms of the detections and non-detections as a function of
angular size.  Since many objects are non-circular, we have simply adopted
the average of the ranges shown in Table 2 as being representative for this
purpose.  There is some indication that detections tend to be smaller diameter 
than non-detections, but the effect is not large.  In particular, in the
1\arcmin\ - 4\arcmin\ range, there are 11 detections and 15 non-detections.
Angular size does not appear to be a primary driver for FUV detectability.

However, there may be some effect at the smallest diameters.  In Figure 26,
we show the observed F($\lambda$1032) from Table 3 against angular size for
the detected SNRs. SNR 0535-660 (N63A) is exceptional, being due to an encounter
of a fast shock with a dense cloud (e.g. Warren et al. 2003), and would 
be located at 45 on the vertical scale of this plot.  However, the Figure
shows the next three brightest objects all below 1.5\arcmin\ in angular size.
Between 1.5\arcmin\ and 4\arcmin, there is significant dispersion in values, but
all objects are as bright or brighter than the single detected larger object.
A single upper limit is shown at the position of the one SNR detected only
in C~III.  However, by referring back to Figure 25, there are numerous other SNRs
in the mid-range of angular diameter that would provide upper limits below
$\rm 1.0 \times 10^{-14} ~ \FLUX$, indicating an even larger dispersion than
shown in Figure 26.

There are at least two objects in the non-detection list with very faint 
optical emission but well-formed X-ray shells (SNR 0453-685, Figure 19 
fourth panel; SNR 0534-699, Figure 23, top panel). These two objects
are modest in size, and may be dominated by non-radiative shocks (e.g.,
Ghavamian et al. 2006 and references therein).  Since
most of our detections show both O~VI and C~III lines at relative strengths
consistent with radiative shocks (Hartigan et al. 1987), it may be 
systematically easier to detect such objects over non-radiative cases.

Another factor in non-detections could simply be foreground extinction.
Accurate optical color excesses or columns densities are not available
for many of the objects we have observed. While the foreground galactic
extinction toward the MCs is generally low, some objects could be
impacted by local extinction.  MC extinction curves tend to show sharper
upturns at short UV wavelengths than the galactic curve (e.g.
Pr\'{e}vot et al. 1984; Fitzpatrick 1985)
so this effect cannot be ignored.  Since many of the detections 
are quite faint, some non-detections could just be due to excessive 
 but unknown attenuation.
%FUV emission is detected even in SNRs where 
%the targeted region exhibits minimal optical and X-ray emission.

With the line flux information in Table 3, we can in principle 
calculate O~VI luminosities for detected objects. However, a number
of assumptions must be made, with the quality of those assumptions
varying considerably from object to object.  For instance, to scale 
an O~VI flux measured through the LWRS aperture to a total O~VI flux,
one must assume a uniform flux across the object and scale to its
projected area, an assumption that may be particularly inaccurate
for the larger SNRs.  Also, since many of the O~VI line profiles
show evidence of overlying absorption, a correction for the missing
flux must be included. A moderately good assumption involves assuming
an optically thin 2:1 ratio for $\lambda$1032:$\lambda$1038 to get a
total O~VI flux.  Finally a correction for overlying extinction
is required to estimate an intrinsic flux from the observed (corrected)
value.  Since extinction values are not known accurately for the individual
objects, we choose below to assume a standard value of E(B - V) = 0.1
and a galactic extinction curve (e.g., Cardelli, Clayton, \& Mathis
1989), implying a factor of 3.5 correction at O~VI.  (Note: this
ignores any intrinsic LMC absorption, so larger corrections may
be appropriate in some cases.)  Proper scaling by distance then 
provides an estimate of the O~VI luminosity of an object.
Rather than attempt this calculation for all detected objects, we
choose to perform it for a few selected objects
where it appears reasonable combinations of assumptions can be
made to provide some insight into the importance of O~VI as a
source of cooling relative to, for instance, the soft X-ray
emission.  

SNR 0454-665 (N11L; Figure 5) is a relatively small diameter SNR with
a well-detected O~VI line. Using the angular size listed in Table 2,
a modest geometric correction factor (3.1) is needed to scale up to
the whole object. If we assume from the line profile that roughly half
of the actual O~VI $\lambda$1032 line is seen through the overlying
absorption, a factor of 1.5 to include $\lambda$1038, the factor of 
3.5 extinction correction factor discussed above, and a distance of 
50 kpc, we derive $\rm L(O~VI) ~=~ 4.2 \times 10^{35}~ergs ~ s^{-1}$.
Williams et al. (1999b) used ROSAT data to estimate 
$\rm L_{x}(0.5 - 2~ keV) ~=~ 8.8 \times 10^{34}~ergs ~ s^{-1}$, or
nearly a factor of 5 less than our estimated L(O~VI).

As a slightly larger object, SNR 0532-710 (N206, Figure 12) has a 
geometric scaling factor from LWRS of $\sim$27. However, the sampled 
region appears to be representative of the shell interior, and if
anything the O~VI emission might be expected to limb-brightened.
From inspection of the line profile, we again estimate $\sim$50\%
attenuation from overlying MC and MW absorption and assume the same 
$\lambda$1038 and extinction correction factors and distance as above,
yielding $\rm L(O~VI) ~=~ 2.4 \times 10^{35}~ergs ~ s^{-1}$.
Williams et al. (2005) report a Chandra/XMM-Newton estimate
of $\rm L_{x}(0.3 - 8~ keV) ~=~ 8(\pm4) \times 10^{35}~ergs ~ s^{-1}$, 
dependent on model spectrum assumptions, or roughly several times
the estimated O~VI luminosity in this case.

By way of contrast, we select an FUV non-detected object, SNR 0453-685 
(D90403; Fig. 19, bottom panel) with a night observing time of 3.8 ks.
Interestingly, this object is quite bright in X-rays but shows no
obvious optical emission in the MCELS data.
We determine a 3$\sigma$ upper limit of $\rm 2.5 \times 10^{-13} ~\FLUX$
in $\lambda$1032. (This assumes an unresolved line and a filled LWRS
aperture.)  With a geometric correction factor of $\sim$16 and
other parameters as for N206 above,
we find $\rm L(O~VI) ~\le~ 1.9 \times 10^{34}~ergs ~ s^{-1}$.
We are unaware of a detailed model of the X-ray emission from this
SNR, but if we simply use the relative X-ray surface brightnesses
for this object and N206 from Williams et al. (1999a, Table 2), 
scale by the projected areas, and assume the same spectral modelling
as done for N206 (Williams et al. 2005), we derive an approximate
X-ray luminosity of 
$\rm L_{x}(0.3 - 8~ keV) ~\simeq~ 6 \times 10^{36}~ergs ~ s^{-1}$,
or at least two orders of magnitude brighter than the O~VI
upper limit. It is unlikely that extinction alone could account
for such a large depression of the L(O~VI), and substantial real
variations of the relative luminosities are likely present. 
While none of these estimates is high fidelity, clearly there is a
broad range of L(O~VI):$\rm L_x$ in the sampled objects, and 
the luminosity in just the O~VI lines can rival the entire soft
X-ray luminosity in some cases.

\section{Summary and Conclusions}

It is clear from the $\fuse$ survey that there are numerous SNRs in
the Magellanic Clouds that are observable in the far ultraviolet. We
have detected 15 SNRs (14 previously undetected in the ultraviolet
plus 0535-660 [N63A]) out of a total of 39 observed, bringing the 
total number of FUV-detected SNRs in the
Magellanic Clouds to 22.  The detected objects span a wide range
of parameter space, from relatively small, bright objects to large (old)
shells, and from both X-ray shells and filled-center morphologies.
Almost as interesting are the non-detections, because the sensitivity
of $\fuse$ allows significant upper limits to be placed in many cases.
These objects must have substantially higher foreground columns and/or
much slower shock velocities at the observed locations than the 
detected objects in order to escape detection.  
Given the likelihood of non-uniform spatial distributions of the FUV
emission, non-detections at the observed positions do not necessarily 
indicate these objects have no FUV emission. Potential clumpiness, 
especially for the largest
angular size SNRs, may be a significant reason for non-detection.
A significant subset of our spectra were contaminated
by stellar light, making the upper limits somewhat less
conclusive in these cases.

The line profiles for detected objects generally show evidence of
significant overlying self-absorption by the ISM of the
host galaxy (if not also the Milky Way, depending on the SNR intrinsic 
line widths).  
Although we typically looked on a sight line through the center of
each SNR, we tend to see emission at wavelengths just redward of
the host galaxy rest frame, as if we are systematically seeing the
receding side of each detected SNR.  This is also indicative of overlying
absorption, which would tend to impact emission from the approaching 
side of each SNR more severely.  In one case, SNR 0536-706, we may
be seeing front-side/back-side shell emission, although the observed
profile in O~VI  could also represent a broader emission line with the 
center removed by O~VI self-absorption (see sec. 5.14).

All but one of the detected SNRs were seen in O~VI $\lambda$1032.  Many 
but not all were detected in C~III $\lambda$977 as well.  A single
SNR was seen marginally in C~III but {\it not} in O~VI.  This may be due to a 
selection effect arising from SNRs with slower shocks (hence, no O~VI)
not being visible through the more substantial ISM absorption at C~III.
However, without additional observations, we cannot rule out the idea 
that O~VI emission is present but patchy in this object, and hence simply 
not seen.

Although many assumptions are involved, estimates of the luminosities 
of these objects in O~VI show a broad range relative to the soft
X-ray luminosities (which also involve a number of assumptions). 
We have examples where the O~VI luminosity is considerably brighter 
than the inferred soft X-ray luminosity, to some of the FUV non-detected
objects where the inferred upper limit on O~VI is
perhaps two orders of magnitude or more below the X-ray luminosity.
The reason or reasons for such diversity are not entirely clear,
although many important parameters, such as line of sight extinction
and clumpiness of the FUV emission,
are not currently well established for many of these objects.

%Some of the detected objects are very faint, but others are moderately bright
%and deserving of follow-up observations. It is ironic that the objects
%now detected with $\fuse$ in the 900 - 1200 \AA\ range have never been 
%observed in the spectral range from 1200 \AA\ up to the atmospheric 
%cutoff, and are unlikely to be observed in this range in the foreseeable 
%future because of the lack of appropriate space-based instrumentation.

\acknowledgments

The $\fuse$ data reported in this paper were obtained with the
NASA-CNES-CSA $\fuse$ satellite, which is operated by Johns Hopkins 
University with financial support through NASA contract NAS 5-32985.
It is a pleasure to thank the $\fuse$ operations team 
for their efforts in obtaining these data.  We also thank Chris 
Smith, Roger Leiton Thompson, and Claudio Aguilera of the MCELS team for 
providing us with the optical images used in this paper. 
This work has been supported by NASA guest investigator grants 
NAG5-12423 and NNG04GJ25G, both to the Johns Hopkins University.

\pagebreak
\bigskip

\section{Appendix}

In this Appendix, we provide brief explanatory comments on each of the
detected Magellanic Cloud SNRs. 

\subsection{SNR 0448-669 (Figure 3)}

This SNR was only recently identified, as part of ongoing analysis of the
MCELS imaging survey (Smith et al. 2004).  The position provided was apparently
poorly centered and instead sits on the northern limb of the optical shell.
No X-ray image is available.  The $\fuse$ detection of
O~VI is modest, although both lines of the doublet are present and
some indication of self-absorption by LMC halo O~VI is evident, especially 
for $\lambda$1032.  The C~III line is not detected with certainty 
although only 3.8 ks of the 15.7 ks integration was in orbital night.
With a line width of nearly 700 $\VEL$, the blue wing of $\lambda$1032 may
be getting clipped by Milky Way halo absorption.

\subsection{SNR 0450-709 (Figure 2)}

This low surface brightness, non-descript shell SNR is one of the largest 
Magellanic SNRs in angular size, first reported by Mathewson et al.
(1985) in the optical and radio.  Williams et al. (2004) have recently provided
a detailed investigation.  The XMM-Newton data show only a faint 
enhancement above background levels with no distinct shape seen.  The X-rays 
appear to be brighter interior to the shell, making the SNR similar to
a handful of large old galactic SNRs such as W28 and 3C400.2 (Long et al.
1991).  Despite its apparent late stage of evolution, the SNR was 
detected in both C~III and O~VI with $\fuse$, although 
as a faint source.  The detection has been aided by the relatively long 
(9.5 ks) orbital night integration for this object. The central velocity 
of the detected SNR emission 
appears to be just slightly redward of the LMC mean velocity. 
The C~III emission appears truncated on the blue side by LMC absorption and is
artificially narrow.  Both O~VI lines appear to be broader lines impacted by 
overlying (mainly LMC) absorption.  Intrinsic fluxes are correspondingly 
uncertain.

\subsection{SNR 0454.5-6713 (Figure 4)}

This SNR was discovered as a diffuse X-ray source, and was found optically
by Smith et al. (1994).  It is located within the LMC H~II region N9.
It's optical morphology is nondescript, with just a few radiative filaments
embedded within faint, diffuse emission.  Faint, diffuse X-ray emission 
fills the region of optical emission and extends somewhat farther to the 
south and southwest.  The Chandra X-ray morphology gives the impression that 
this SNR is a member of the composite or ``mixed morphology" class, with 
a faint outer shell and bright, filled center.  
The bright central X-ray emission in the Chandra image is 
probably due to ejecta, and Seward et al. (2004) point to a likely Type Ia
supernova origin.  The $\fuse$ aperture position, based on the estimated
optical centroid, is clearly off-center based on the X-ray image and
samples primarily the diffuse regions in X-ray and optical.  
The strong, relatively narrow C~III and O~VI emission lines lie just longward
of the LMC rest frame velocity and appear to be due to radiative shocks. 
The lines could be impacted by blue wing absorption, but the lack of an
extended red wing on the strong, narrow lines argues for relatively
low velocity dispersion within the aperture.

\subsection{SNR 0454-665 (N11L; Figure 5)}

This SNR is a member of the original group of SNRs identified in the LMC 
by Mathewson \& Clarke (1973) using the optical [S~II] to H$\alpha$ 
criterion, and has been studied in detail by Williams et al. (1999b). 
Although it is part of the large
N11 H~II region complex, it forms a well-defined and relatively isolated
shell with an apparent ``jet" or breakout to the NE, reminiscent of
the optical jet in Cas A (Fesen \& Gunderson 1996). With a relatively 
small angular size,
the ROSAT PSPC image shows little structure and leaves the impression 
of filled-center emission.
The $\fuse$ spectrum with only 2.2 ks of orbital night nonetheless shows 
very bright and well-detected O~VI $\lambda\lambda$1032,1038
and C~III lines with evidence for absorption on the blue sides
of the profiles. In particular, the $\lambda$1032 line provides evidence 
of a much broader and stronger line than is actually detected.
Radiative shock emission apparently dominates the line emission from
material in the $\fuse$ aperture.

\subsection{SNR 0506-680 (N23; Figure 6)}

Once again, the faintness of the diffuse, patchy optical emission from this
SNR has caused the Mathewson et al. (1983) coordinate to be off-center
in comparison with the Chandra X-ray data, which extends considerably to 
the north and west from the bulk of the optical emission.  In this case, the 
$\fuse$ aperture lies on the SE bright X-ray limb.  Even with only 2.7 ks of orbital night integration, both O~VI and C~III are detected, although the C~III
line is quite faint.  The O~VI emission is quite strong and very broad 
relative to C~III in this object. 
The line shape for O~VI $\lambda$1032 makes it clear that both 
galactic and LMC overlying absorption are impacting the observed fluxes.  
The brightness of the X-ray and O~VI intensity and the faintness of the
optical emission and C~III line make it likely that non-radiative shocks
contribute significantly to the observed emission.

\subsection{SNR 0509-687 (N103B; Figure 7)}

This young, small diameter SNR nearly fits within the nominal $\fuse$ 
LiF1A LWRS aperture. The bright X-ray emission corresponds to the
region of knotty optical emission, implicating a shock-cloud interaction 
on the west side, but fainter X-ray emission extends toward the east. 
No discernable C~III emission is present, although the small angular 
size of the SNR makes it possible in principle that the $\fuse$ SiC 
apertures could have been misaligned enough to miss or only partially 
cover the SNR. (We note that the LiF2B channel, which also covers the
O~VI spectral region, was apparently misaligned--see main text.) 
The O~VI $\lambda$1032 emission is quite strong and broad, and the
$\lambda$1038 line is obviously impacted more severely by overlying 
absorption.  A small amount of scattered starlight contaminates the 
spectrum near O~VI, causing the apparent pedestal of emission.

\subsection{SNR 0520-694 (Figure 8)}

This relatively faint and moderately sized optical shell SNR 
lies in a rich star field within the bar of the LMC.  Significant
stellar emission contaminates the $\fuse$ aperture position, but
% ??has a diffuse appearance in X-rays. ??
we feel confident
in claiming detection by virtue of the strong, relatively narrow C~III 
emission just longward of the LMC rest velocity.
Although an O~VI P-Cygni feature in the stellar spectrum cannot
be ruled out without more information, an interpretation of the dip
at $\lambda$1032 as ISM absorption would make it likely that the
apparent emission just longward of the LMC rest velocity is faint,
broad O~VI emission from the SNR. This is the feature measured and reported
in Table 3.

\subsection{SNR 0527-658 (DEML 204; Figure 9)}

This is one of the larger SNRs in the survey, and is an apparent member of
the composite class, with a large outer optical shell and filled-center
X-ray emission.  This is the only SNR in the survey that shows C~III
emission but no detectable emission in O~VI, and even the C~III
detection is somewhat marginal.  However, the appearance
of the nominal C~III line is consistent with the hypothesis we have put
forward in the main text for why there are almost no such detections: the 
intrinsic C~III SNR line is largely absorbed by intervening ISM.  For this
object, the kinematics are such that the C~III line peeks out on the red 
side of the LMC absorption and is just visible. The intrinsic C~III
line is likely quite strong in this large (older) radiative shell-type
SNR, and the absence of detectable O~VI indicates few shocks remaining
at velocities in excess of 160 $\VEL$. There is a possible emission
line at the position of C~II $\lambda$1037.02, the excited state
component of the more well known C~II $\lambda$1036.34 ISM absorption
line, although significant overlying absorption may be present. The 
possible presence of this line is also consistent with the emission
from this object being dominated by slow radiative shocks.

\subsection{SNR 0528-692 (LHG40; Figure 10)}

This SNR was first reported in X-rays by Long, Helfand, \& Grabelsky (1981)
and optical and radio detections were provided by Mathewson et al. (1984).
A faint optical radiative shell is visible, but the X-ray emission is
ill-defined in the ROSAT PSPC data.  This is another object in the LMC 
bar, and stellar contamination is nearly impossible to avoid.
The $\fuse$ spectrum of this object includes only orbital day data,
and so additional lines are due to residual dayglow, including the feature
near Milky Way C~III rest velocity.  Despite the stellar
contamination, the SNR O~VI emission is clearly seen in both lines at 
the LMC velocity.  Interestingly, no C~III at LMC velocities is detected, 
which is surprising given the radiative shell nature of the optical emission.
It may be that the receding shell is too faint to be sween and the near-side 
emission is lost in the ISM absorption from the LMC.
 
\subsection{SNR 0530-701 (Figure 11)}

This SNR was only recently identified, as part of ongoing analysis of the
MCELS imaging survey (Smith et al. 2004), and appears to be a large, old,
radiative shell-type SNR.  No X-ray data are available.  
The supplied coordinate for this object was on the southern rim instead of 
the center.  The $\fuse$ detection of O~VI is quite weak but moderately broad.  
Absorption may affect the blue side of the $\lambda$1032 line, and 
$\lambda$1038 is only marginally present.
The C~III line is not detected with certainty, although the small fraction
of orbital night data for this observation (and the use of the total
data set for this object) both complicate the interpretation of C~III.

\subsection{SNR 0532-710 (N206; Figure 12)}

This composite or ``mixed morphology" SNR shows a classic symmetrical 
shell in the optical (with the
shocked radiative emission showing particularly well in [S~II]),
but a center-filled X-ray morphology.  A detailed study has been published 
by Williams et al. (2005), with the addition of a possible pulsar wind
nebula discovered in X-ray and radio data. In this moderate sized SNR, the
central X-ray emission is likely due to ejecta. The $\fuse$ spectrum shows 
faint, broad O~VI emission and a very narrow C~III line just longward of the
LMC rest velocity.  Clearly overlying absorption is impacting these profiles, 
and the blue side of C~III could just be missing. It is 
likely that the observed emission is due to the radiative shell.

\subsection{SNR 0534-705 (DEM 238; Figure 13)}

This moderately sized faint shell-type optical SNR was reported by Mathewson 
et al. (1983). The Chandra data show extended faint X-ray emission
filling the optical shell, but a brighter central concentration
(somewhat offset toward the southwest) that is likely due to ejecta.
Even though the $\fuse$ aperture is filled partially with ejecta,
the narrowness of the observed line profiles makes it likely that the 
observed lines are fron the readiative shell.
Both O~VI lines are seen but $\lambda$1038 is weak, implying overlying
absorption.  The C~III line is comparable to O~VI $\lambda$1032 in peak 
intensity but is narrower,  largely attributable to ISM absorption on 
the blue side of the profile.

\subsection{SNR 0535-660 (N63A; Figure 14)}

This SNR was detected with IUE (Benvenuti et al. 1980) but has not been
observed with more modern instrumentation.  
The IUE spectra only showed C~IV $\lambda$1550, He~II $\lambda$1640,
and C~III $\lambda$1909 emission lines superimposed on a continuum
of stellar contamination and/or dust-scattered starlight. 
The SNR is buried within an 
extended H~II region and OB association.  The bizarre optical morphology
is dominated by a shocked cloud located on the near side of the SNR
from our sight line.  The true size and shape of the SNR is apparent
from the soft X-rays (e.g., Warren et al. 2003).
The $\fuse$ aperture covered this shocked cloud and the spectrum is 
contaminated by significant continuum.  Nonetheless, very strong
O~VI emission lines are detected just shortward of the LMC rest velocity. 
There is no clear evidence for C~III.  Since IUE observed C~III 
$\lambda$1909, it is likely that
the SiC channels were misaligned by enough that they missed the bright
radiative shocked cloud.  The strength of the O~VI emission implies a
radiative shock origin, even though C~III and other cooler ions
are missing. Given that 
It is probable that the FUV emission is dominated by
radiative shocks. 
This is consistent with a dense, shocked cloud 
being hit relatively recently by the blast wave.

\subsection{SNR 0536-706 (DEML 249; Figure 15)}

This SNR is a large, faint optical shell SNR that is 
brighter on the eastern side of the shell.  The X-ray emission from
ROSAT PSPC is low resolution, but rather diffuse and strongest at the 
center, possibly indicating it is a member
of the mixed-morphology class (Williams et al. 1999a).
The $\fuse$ spectrum shows a narrow C~III line centered at the LMC
rest velocity and clearly double-peaked O~VI lines that bracket
the LMC rest velocity.
The presence of C~III emission at the rest velocity argues that
this object must be on the near side of the LMC, and thus suffers
little overlying absorption from LMC ISM C~III. The O~VI line
morphology could then arise from one of two mechanisms: either
O~VI emission arises from approaching and receding sides of the shell
and we are resolving this structure, or a broader O~VI emission line
is undergoing significant self-absorption from within the SNR itself.
Whatever the cause, this spectral character is unique in the objects 
we have observed in this survey.

\subsection{SNR 0104-723; Figure 16}

This is the only SMC remnant observed to date in the $\fuse$ survey, and 
it was first reported by Mathewson et al. (1984).  It is
detected in O~VI. Both O~VI lines appear to be impacted by overlying
absorption from both our Galaxy and the SMC. C~III may be marginally
detected.  The strong emission at $\lambda$977 in the rest frame is
backscattered solar emission in the $\fuse$ SiC channel and is not real.

%%%%%%%%%%%%%  FIGURES  %%%%%%%%%%%%%%%%%%%%%%%%%%%%%%%%%%%%%

\pagebreak
\bigskip
\centerline{\bf Figure Captions}
\bigskip
% 
% Figure captions:

Note: Due to size restrictions, all figures included as 
separate `JPG' files.

% FIGURE 1:
{\bf Fig.~1:} Overview image of the Large Magellanic Cloud showing the 
distribution of detected and non-detected SNRs. Detected objects are
indicated with stars symbols, with filled stars being previous
detections (from Table 1) and white stars being new detections from 
Table 2.  Non-detections are indicated with crosses.
H$\alpha$ image from Gaustad et al. (2001).

% FIGURE 2-16:
\noindent
{\bf Fig.~2:} Summary figure for SNR 0450-709. 
The top panel shows optical and X-ray images 
of the SNR with the $\fuse$ LWRS aperture (30\arcsec\ square) overlaid 
for the location and position angle at the time of the observation. 
Top left: H$\alpha$; Top middle: [S~II]; 
Top right: soft X-ray (from the source indicated by the label).  
The $\fuse$ aperture (30\arcsec\ square) provides the scale.
The middle and bottom panels show $\fuse$ data sections centered on
C~III $\lambda$977, and O~VI $\lambda\lambda$1032,1038, respectively.
Each spectral panel contains two horizontal bars with tick marks,
indicating the positions of potential overlying absorption features,
one at zero velocity for Galactic absorption (labelled MW), and one
shifted to the mean velocity of the host Magellanic Cloud (+275
$\VEL$ for LMC; +165 $\VEL$ for SMC). The long, bold tick marks are
at the expected positions of the C~III and O~VI resonance lines,
which are in emission from the SNR, but potentially in absorption
by overlying gas.  Medium tick marks indicate positions of $\rm H_2$
transitions, and short tick marks indicate positions of interstellar
O~I lines. Very short ticks in the bottom panel indicate the
position of C~II $\lambda$1036.3, another very strong ISM absorption line.
The earth symbol indicates terrestrial H airglow emissions from Ly
$\gamma$ (C~III panel) and Ly $\beta$ (O~VI panel).

\noindent
{\bf Fig. 3:} Same as Figure 2, but for SNR 0448-669.
In this case, no X-ray data are available, and the coordinate
provided was apparently off-center and on the northern limb
of the shell.

\noindent
{\bf Fig. 4:} Same as Figure 2, but for SNR 0454.5-6713.

\noindent
{\bf Fig. 5:} Same as Figure 2, but for SNR 0454-665 (N11L).

\noindent
{\bf Fig. 6:} Same as Figure 2, but for SNR 0506-680.

\noindent
{\bf Fig. 7:} Same as Figure 2, but for SNR 0509-687 (N103B).

\noindent
{\bf Fig. 8:} Same as Figure 2, but for SNR 0520-694.
In this case, no X-ray data are available.

\noindent
{\bf Fig. 9:} Same as Figure 2, but for SNR 0527-658.

\noindent
{\bf Fig. 10:} Same as Figure 2, but for SNR 0528-692.

\noindent
{\bf Fig. 11:} Same as Figure 2, but for SNR 0530-701.
In this case, no X-ray data are available, and the coordinate
provided was apparently off-center and on the southern limb
of the shell. Due to the small fraction of orbital night data,
the total dat set is used for this figure.  The additional lines
visible are due to dayglow emissions, including especially the
strong line at Milky Way C~III rest velocity.

\noindent
{\bf Fig. 12:} Same as Figure 2, but for SNR 0532-710.

\noindent
{\bf Fig. 13:} Same as Figure 2, but for SNR 0534-705.

\noindent
{\bf Fig. 14:} Same as Figure 2, but for SNR 0535-660 (N63A).
The aperture includes the bright optical shocked cloud, which is
known to be on the near-side (blue-shifted) of the SNR. Bright
O~VI emission is seen (despite considerable stellar contamination)
but there is no indication of C~III.

\noindent
{\bf Fig. 15:} Same as Figure 2, but for SNR 0536-706.

\noindent
{\bf Fig. 16:} Same as Figure 2, but for SNR 0104-723.
This is the only SMC SNR surveyed to date.  The velocity scale of the top
bar has been adjusted to that of the SMC.

% FIGURE 17:
\noindent
{\bf Fig.~17:} C~III and O~VI spectral regions for the non-detection of
SNR 0507-685 in the LMC. The upper line in each panel shows the total
data set (10.9 ks CHECK), while the lower line shows the orbital-night 
data section for comparison (4.9 ks CHECK). The dashed lines show the zero
point of the total data sets, which have been arbitrarily offset upwards
for display purposes.  The apparent weak emission at the LMC C~III
position in the total data is absent in the night-only segment, indicating
it is not intrinsic SNR emission.

% FIGURE 18:
\noindent
{\bf Fig.~18:}  Same as Figure 18, but for SNR 0519-697, an example where 
stellar light contaminated the $\fuse$ data. It is conceivable that
LMC ISM absorption is present but being filled in with emission
from the SNR.  However, without a separate template spectrum of the
star, there is no way to tell.  Hence, the ability to place an
upper limit on emission is compromised in cases such as this.

% FIGURE 19-24:
\noindent
{\bf Fig.~19:} Optical and X-ray images of SNRs declared 
non-detections, in the same format as the top panels in Figures
2-16.  The nominal $\fuse$ LiF1 LWRS aperture positions are 
superimposed and provide the scale for each panel.  Shown are
SNRs 0449-693, 0453.1-6655, 0453-672, and 0453-685.

\noindent
{\bf Fig.~20:} Same as Figure 19, but for SNRs 0455-687 (N86), 
0500-702, 0506-657, and RXJ0507-685
(see also Figure 17). 

\noindent
{\bf Fig.~21:} Same as Figure 19, but for SNRs 0513-692, 0519-697
(see also Fig. 18),  0521-657, and 0523-679.

\noindent
{\bf Fig.~22:} Same as Figure 19, but for SNRs 0524-664, 0528-672,
0532-673, and 0532-675.

\noindent
{\bf Fig.~23:} Same as Figure 19, but for SNRs 0534-699, 0536-676, 
0536-692, and 0536-693.

\noindent
{\bf Fig.~24:} Same as Figure 19, but for SNRs 0538-693, 
0543-689, and 0547-697, where both ``a" and ``b" components 
(DEML316A and B) were observed separately. (See Williams et al. 1997.)

\noindent
{\bf Fig.~25:} Histograms of detected SNRs (solid line) and non-detected SNRs
(dashed line) as a function of angular size (see text).  The bin centered at 
5.5 represents all objects larger than 5\arcmin.

\noindent
{\bf Fig.~26:} A plot of measured O~VI $\lambda$1032 flux versus angular size
for detected SNRs. There is some tendancy for the smallest objects to have
the highest O~VI fluxes, but a large dispersion of values is indicated for
objects above 1.5\arcmin\ in size.  SNR 0535-660 (N63A) would be at 45 units
on the vertical scale.  The single detection in C~III-only, SNR 0527-658, is
shown as an upper limit, but other non-detections would also produce upper
limits at or below the limit on this object.

%\clearpage
%\begin{figure}
%\plotone{fig1.eps}
%%\caption{\label{fig:fig1}} 
%\end{figure}
%
%% Figs 2-16 are hits figures
%
%\clearpage
%\begin{figure}
%\plotone{fig2_5.eps}
%%\plotone{0450_bw.eps}
%%\caption{\label{fig:fig2}} 
%\end{figure}
%
%\clearpage
%\begin{figure}
%%\plotone{0448_bw.eps}
%\plotone{fig6_9.eps}
%%\caption{\label{fig:fig3}} 
%\end{figure}
%
%\clearpage
%\begin{figure}
%\plotone{fig10_13.eps}
%%\plotone{0454.5_bw.eps}
%%\caption{\label{fig:fig4} }
%\end{figure}
%
%\clearpage
%\begin{figure}
%%\plotone{0454.eps}
%\plotone{fig14_16.eps}
%%\caption{\label{fig:fig5}} 
%\end{figure}
%
%\clearpage
%\begin{figure}
%\epsscale{0.5}
%%\plotone{0454.eps}
%\plotone{fig17_18.eps}
%%\caption{\label{fig:fig5}} 
%\end{figure}
%
%\clearpage
%\begin{figure}
%%\plotone{0454.eps}
%\plotone{fig19_22.eps}
%%\caption{\label{fig:fig5}} 
%\end{figure}
%
%\clearpage
%\begin{figure}
%%\plotone{0454.eps}
%\plotone{fig23_26.eps}
%%\caption{\label{fig:fig5}} 
%\end{figure}
%
%%%%%%%%%%TABLES%%%%%%%%%%%%%%%%%%%%

%%%%% Table 1

\begin{deluxetable}{lccccc}
\tablenum{1}
\tablewidth{0pt}
\label{tbl:table1}
\tablecaption{Previous $\fuse$ Observations of Magellanic Cloud SNRs}
\tablehead{
\colhead{Object} &
\colhead{RA(J2000)} &
\colhead{Dec(J2000)} &
\colhead{ProgID} &
\colhead{Detection?} &
\colhead{Reference}
%\colhead{Reference\tablenote{a}}
}
\startdata
SNR0057-7226        & 00:59:27   & $-$72:10:05  & P103,P203 & Yes  & 1 \\
SNR0102-7219        & 01:04:04   & $-$72:01:50  & A075,C083 & Yes  & 2 \\
SNR0505-679 (DEM L71) & 05:05:43  & $-$67:52:38  & P214,C072 & Yes  & 3 \\
SNR0509-675         & 05:09:32   & $-$67:31:17  & P214     & Yes   & 3  \\
SNR0519-690         & 05:19:34   & $-$69:02:10  & P214     & Yes  & 3  \\
SNR0525-696 (N132D) & 05:25:01   & $-$69:38:34  & A075,B095 & Yes  & 4 \\
SNR0525-661 (N49)   & 05:26:04   & $-$66:05:18  & X005,C055 & Yes  & 5,6 \\
SNR0548-704         & 05:47:50   & $-$70:24:52  & P214     & No   & 3  \\
\enddata

References: 
(1) Danforth et al. (2003);
(2) Sasaki et al. (2006);
(3) Ghavamian et al. (2006);
(4) Beasley et al. (2006); 
(5) Blair et al. (2000b); 
(6) Sankrit et al. (2004). 

%\tablenotetext{a}{(1) Downes (1986). (2) This paper.}

\end{deluxetable}

%%%% Table 2
\clearpage

\input{tab2.tex}

\clearpage

%%%%% Table 3

\clearpage
%\small

\begin{deluxetable}{lccc}
\tablenum{3}
\tabletypesize{\small}
\tablewidth{0pt}
\label{tbl:table3}
\tablecaption{Measured Parameters for Magellanic Cloud SNR Spectra}
\tablehead{
\colhead{Object} &
\colhead{O~VI$\lambda1032$} &
\colhead{O~VI$\lambda1038$} &
\colhead{C~III$\lambda977$} \\
\colhead{~~~~Parameter} &
\colhead{} &
\colhead{} &
\colhead{} 
%\colhead{Reference\tablenote{a}}
}
\startdata
SNR0448-669 (E90001)      &          &         &   \\
~~~$\lambda$(cent) [\AA]  & 1033.18  & 1038.87 & ... \\
~~~V(cent) [km/s]         &  +369    & +361    & ...  \\
~~~$\Delta$V(FWZI) [km/s] & 676      & ...     & ... \\
~~~F($\lambda$) [$\FLUX$] & 2.8E-14  & 9.4E-15 & ...  \\
%                          &          &         &   \\
SNR0450-709 (D90401)      &          &         &   \\
~~~$\lambda$(cent) [\AA]  & 1032.62  & 1038.32 & 978.13 \\
~~~V(cent) [km/s]         &  +206    & +201    & +341  \\
~~~$\Delta$V(FWZI) [km/s] & 627      & ...     & 219 \\
~~~F($\lambda$) [$\FLUX$] & 1.2E-14  & 3.6E-15 & 1.2E-14  \\
%                          &          &         &   \\
SNR0454.5-6713 (D90405)      &          &         &   \\
~~~$\lambda$(cent) [\AA]  & 1032.94  & 1038.64 & 978.25 \\
~~~V(cent) [km/s]         &  +299    & +284    & +377  \\
~~~$\Delta$V(FWZI) [km/s] & 336      & ...     & 299 \\
~~~F($\lambda$) [$\FLUX$] & 3.2E-14  & 1.6E-15 & 7.8E-14  \\
%                          &          &         &   \\
SNR0454-665 (D90406)      &          &         &   \\
~~~$\lambda$(cent) [\AA]  & 1032.67  & 1038.40 & 978.28 \\
~~~V(cent) [km/s]         &  +220    & +225    & +387  \\
~~~$\Delta$V(FWZI) [km/s] & 563      & ...     & 291 \\
~~~F($\lambda$) [$\FLUX$] & 1.7E-13  & 6.5E-14 & 1.3E-13  \\
%                          &          &         &   \\
SNR0506-680 (D90409)      &          &         &   \\
~~~$\lambda$(cent) [\AA]  & 1032.94  & 1038.64 & 978.34 \\
~~~V(cent) [km/s]         &  +294    & +292    & +405  \\
~~~$\Delta$V(FWZI) [km/s] & 684      & ...     & 251 \\
~~~F($\lambda$) [$\FLUX$] & 1.4E-13  & 3.1E-14 & 2.3E-14  \\
%                          &          &         &   \\
SNR0509-687 (D90410)      &          &         &   \\
~~~$\lambda$(cent) [\AA]  & 1033.01  & 1039.11 & ... \\
~~~V(cent) [km/s]         &  +320    & +431    & ...  \\
~~~$\Delta$V(FWZI) [km/s] & 533      & ...     & ... \\
~~~F($\lambda$) [$\FLUX$] & 1.6E-13  & 6.9E-14 & $<$1.0E-14  \\
                          &          &         &   \\
                          &          &         &   \\
SNR0520-694 (D90448)      &          &         &   \\
~~~$\lambda$(cent) [\AA]  & 1033.41:  &  ...    & 978.45 \\
~~~V(cent) [km/s]         &  +436:   &  ...    & +439  \\
~~~$\Delta$V(FWZI) [km/s] & 490:      & ...     & 393 \\
~~~F($\lambda$) [$\FLUX$] & 6.7E-14:  & ...    & 2.0E-13  \\
%                          &          &         &   \\
SNR0527-658 (D90417)      &          &         &   \\
~~~$\lambda$(cent) [\AA]  & ...  & ... & 978.38 \\
~~~V(cent) [km/s]         &  ...    & ...    & +418  \\
~~~$\Delta$V(FWZI) [km/s] & ...     & ...     & 315 \\
~~~F($\lambda$) [$\FLUX$] & $<$1.0E-14  &  ...   & 2.2E-14  \\
%                          &          &         &   \\
SNR0528-692 (D90418)      &          &         &   \\
~~~$\lambda$(cent) [\AA]  & 1033.00  & 1038.76 & ... \\
~~~V(cent) [km/s]         &  +320    & +330    & ...  \\
~~~$\Delta$V(FWZI) [km/s] & 453      & ...     & ... \\
~~~F($\lambda$) [$\FLUX$] & 4.8E-14  & 3.5E-14 & ...  \\
%                          &          &         &   \\
SNR0530-701 (E90005)      &          &         &   \\
~~~$\lambda$(cent) [\AA]  & 1032.97  & 1038.69 & ... \\
~~~V(cent) [km/s]         &  +308    & +309    & ...  \\
~~~$\Delta$V(FWZI) [km/s] & 604      & ...     & ... \\
~~~F($\lambda$) [$\FLUX$] & 1.4E-14  & 3.0E-15 & ...  \\
%                          &          &         &   \\
SNR0532-710 (D90419)      &          &         &   \\
~~~$\lambda$(cent) [\AA]  & 1033.13  & 1038.90 & 978.38 \\
~~~V(cent) [km/s]         &  +355    & +370    & +418  \\
~~~$\Delta$V(FWZI) [km/s] & 650      & ...     & 315 \\
~~~F($\lambda$) [$\FLUX$] & 1.5E-14  & 1.8E-15 & 2.2E-14  \\
%                          &          &         &   \\
SNR0534-705 (D90422)      &          &         &   \\
~~~$\lambda$(cent) [\AA]  & 1033.34  & 1039.04 & 978.42 \\
~~~V(cent) [km/s]         &  +416    & +410    & +430  \\
~~~$\Delta$V(FWZI) [km/s] & 370      & ...     & 231 \\
~~~F($\lambda$) [$\FLUX$] & 6.1E-14  & 2.0E-14 & 2.7E-14  \\
                          &          &         &   \\
                          &          &         &   \\
                          &          &         &   \\
SNR0535-660 (D90423)      &          &         &   \\
~~~$\lambda$(cent) [\AA]  & 1032.67  & 1038.42 & ... \\
~~~V(cent) [km/s]         &  +221    & +231    & ...  \\
~~~$\Delta$V(FWZI) [km/s] & 480      & ...     & ... \\
~~~F($\lambda$) [$\FLUX$] & 4.5E-13  & 3.5E-14 & ...  \\
%                          &          &         &   \\
SNR0536-706 (D90426)      &          &         &   \\
~~~$\lambda$(cent) [\AA]  & 1033.05  & 1038.88 & 978.15 \\
~~~V(cent) [km/s]         &  +331    & +364    & +347  \\
~~~$\Delta$V(FWZI) [km/s] & 608      & ...     & 200 \\
~~~F($\lambda$) [$\FLUX$] & 3.4E-14  & 8.5E-15 & 2.7E-14  \\
%                          &          &         &   \\
SNR0104-723 (D90444)      &          &         &   \\
~~~$\lambda$(cent) [\AA]  & 1032.66  & 1038.35 & 978.34 \\
~~~V(cent) [km/s]         &  +218    & +211    & +405  \\
~~~$\Delta$V(FWZI) [km/s] & 604      & ...     & 290 \\
~~~F($\lambda$) [$\FLUX$] & 2.6E-14  & 1.2E-14 & 1.6E-14  \\
%                          &          &         &   \\
\enddata

%\tablenotetext{a}{(1) Downes (1986). (2) This paper.}

\end{deluxetable}

%%%%%%%%%%%%%%%%%%%%%%%%%%%%%%%%%%%%%%%%%%%%%%%%%%%%%%%%%%%%%%%%%%%%%%%%

\end{document}

%% file: tab2.tex
% include file for paper.
%\documentclass[aaspp4,11pt]{aastex}
%\begin{document}
\begin{deluxetable}{lccccccc}
\tablenum{2}
\tabletypesize{\footnotesize}
\tablecaption{Observation Log for LMC and SMC Remnants Observed with FUSE}
\tablehead{\colhead{FUSE ID} & \colhead{SNR ID} & \colhead{Optical ID\tablenotemark{a}} &
\colhead{$\alpha_{J2000}$} & \colhead{$\delta_{J2000}$}  &
\colhead{Size (\arcsec)\tablenotemark{b}}  &       
\colhead{Exp.(ks)\tablenotemark{e}} & \colhead{Detected?\tablenotemark{f}}  }
\startdata
E90001 & 0448$-$669  &  ...  &   04:48:23.0 & $-$66:58:56.0 & 240$\times$144 &   15.7(3.8)  &  Yes\\
E90002 & 0449$-$693  &  ...  &   04:49:40.0 & $-$69:21:49.0 & 120 &   11.9(3.4)  &  No\tablenotemark{g}\\
D90401 & 0450$-$709  &  ...  &  04:50:29.7  &  $-$70:50:25.7  &  390$\times$281  &  22(9.5)  &  Yes\\
D90402 & 0453.1$-$665  & (In)N4D  &  04:53:14.0  &  $-$66:55:10.0  &  320$ \times$200  & 18.8(9.4)  & No*\\
D90403 & 0453$-$685  &  ...  &  04:53:36.9  &  $-$68:29:29.0  &  140$\times$131  &  9.3(3.7)  &  No*\\
D90404 & 0453$-$672  &  DEML25,N185  &  04:53:47.4  &  $-$69:59:15.0  &  360   &  23.0(8.4)  &  No*\\
D90405 & 0454.5$-$6713  &  In(N9)  &  04:54:33.0  &  $-$67:12:50.0  &  294$ \times$156 &  18.2(8.1)  &  Yes\\
D90406 & 0454$-$665  &  N11L  &  04:54:48.5  &  $-$66:25:40.5  &  60$\times$53  &  15.3(2.2) &  Yes\\
D90445 & 0455$-$687  &  N86  &  04:55:42.1  &  $-$68:39:14.9  &  360$\times$183  &  10.9(7.3)  &  No\\
D90446 & 0500$-$702  &  N186D  &  04:59:54.7  &  $-$70:08:07.1  &  116  &  11.3(3.6)  &  No*\\
D90409 & 0506$-$680  &  N23  &  05:05:55.4  &  $-$68:01:57.0  &  93$\times$84   &  11.6(2.7)  &  Yes\\
E90003 & 0506$-$657  &  ...  &   05:06:05.0 & $-$65:41:30.0 & 290 &   12.8(3.9)  &  No\\
D90410 & 0509$-$687  &  N103B  &  05:08:59.1  &  $-$68:43:34.2  &  39$\times$42  &  22.5(6.0)  &  Yes\\
D90411 & 0507$-$685  &  RXJ0507-68  &  05:07:30.0  &  $-$68:47:00.0  &  12(450)\tablenotemark{h}  &   9.9(4.6)  &  No\\
D90412 & 0513$-$692  &  ...  &  05:13:41.8  &  $-$66:22:53.7  &  225$\times$191  &  5.9(4.9)  &  No\\
D90447 & 0519$-$697  &  N120  &  05:18:44.2  &  $-$69:39:12.4  &  104  &  20.5(8.5)  &  No*\\
D90448 & 0520$-$694  &  ...  &  05:19:45.1  &  $-$69:25:58.7  &  138$\times$104  &  21.5(7.2)  &  Yes\\
E90004 & 0521$-$657  &  ...  &   05:21:44.0 & $-$65:42:05.0 & 170$\times$110 &   15.6(5.1)  &  No\\
D90415 & 0523$-$679  &  N44-shell3  &  05:23:05.0  &  $-$67:52:30.0  &  180   &  13.0(4.5)  &  No\\
D90416 & 0524$-$664  &  DEML175A  &  05:24:20.5  &  $-$66:23:39.6  &  180$ \times$165  &  8.7(2.9)  &  No\\
D90418 & 0528$-$692  &  LHG40  &  05:27:39.3  &  $-$69:12:14.8  &  143  &   6.6(0)  &  Yes\\
D90417 & 0527$-$658  &  DEML204  &  05:27:57.1  &  $-$65:50:11.1  &  237$\times$212  &  11.5(6.9)  &  Yes\tablenotemark{i} \\
D90432 & 0528$-$672\tablenotemark{c}  &  ...  &  05:28:25.8  &  $-$67:14:42.5   &  675  &  21.8(10.6)  &  No\\
E90005 & 0530$-$701  &  ...  &   05:30:37.0 & $-$70:08:40.0 & 145 &  11.9(2.8)  &  Yes\\
D90419 & 0532$-$710  &  N206  &  05:31:55.9  &  $-$71:00:14.0  &  222$\times$210  &  16.3(4.8) &  Yes\\
D90420 & 0532$-$675\tablenotemark{c}  &  LHG48  &  05:32:23.0  &  $-$67:31:02.0  &  790: &  10.5(6.2)  &  No\\
D90433 & 0532$-$673\tablenotemark{c}  &  LHG48  &  05:32:23.2  &  $-$67:31:59.8  &  790: &  14.4(1.4)  &  No\\
D90421 & 0534$-$699  &  ...  &  05:33:60.0   &  $-$69:54:58.6  &  144$\times$138  &  10.9(1.5)  &  No\\
D90422 & 0534$-$705  &  DEML238  &  05:34:17.4  &  $-$70:33:25.1  &  179$\times$162  &  11.4(4.3)  &  Yes\\
D90423 & 0535$-$660  &  N63A  &  05:35:43.1  &  $-$66:02:05.9  &  68  &  13.4(5.0)  &  Yes\\
D90449 & 0536$-$693  &  Honeycomb  &  05:35:49.6  &  $-$69:18:16.0  &  120$\times$60  &  19.2(8.3)  &  No*\\
D90425 & 0536$-$676  &  DEML241,  &  05:36:01.5  &  $-$67:34:49.6  &   188  &  12.6(2.0)  &  No\\
 &   &  LHG60  &   &    &     &    &  \\
D90426 & 0536$-$706  &  DEML249  &  05:36:07.4  &  $-$70:38:47.1  &  162$\times$129  &  8.3(3.1)  &  Yes\\
D90431 & 0536$-$692  &  LHG62  &  05:36:07.7  &  $-$69:11:52.6  &  469$ \times$375  &  13.4(3.0)  &  No\\
D90427 & 0538$-$693  &  ...  &  05:38:14.0  &  $-$69:21:36.8  &  169  &  18.4(0.5)   &  No*\\
D90428 & 0543$-$689  &  DEML299  &  05:43:07.2  &  $-$68:58:52.0  &  292$\times$243  &  10.1(5.6)  &  No*\\
D90429 & 0547$-$697a  &  DEML316A  &  05:47:02.0  &  $-$69:41:18.0  &  102  &  12.8(4.3)  &  No\\
D90430 & 0547$-$697b  &  DEML316B  &  05:47:00.0  &  $-$69:43:00.0  &  190$\times$150  &  9.8(3.6)  &  No\\
D90444 & 0104$-$723  &  IKT25\tablenotemark{d}  &  01:06:18.9  &  $-$72:05:37.4  &  97  &  7.6(0)  &  Yes\\
\enddata
\tablenotetext{a}{Optical IDs (DEM) from Davies, Elliot, \& Meaburn  
(1976), (LGH) from Long, Helfand, \& Grabelsky (1981), and (N) from Henize (1956).}
\tablenotetext{b}{Diameters and center coordinates from SNR catalogs
of Mathewson et al. (1983, 1984, 1985)  or Williams et al. (1999) in most cases. A few obviously incorrect 
values have been updated from inspection of the images provided in this work.}
\tablenotetext{c}{Very large SNR listed by Mathewson et al. (1985).  Object was observed at the Mathewson et al.
coordinate (second listing) and at a position where optical echelle spectra had showed high velocity (Danforth 2003; first
listing. The SNR was undetected at both locations.}
\tablenotetext{d}{SNR ID from X-ray catalog of Inoue, Koyama, \&  
Tanaka (1983). }
\tablenotetext{e}{Total integration times for each object. The orbital night
time fractions are given in parentheses. Note two objects with no orbital night time
were still detectable.} 
\tablenotetext{f}{A ``No" with an asterisk indicates those objects for which
stellar contamination affects the ability to see potential faint emission.}
\tablenotetext{g}{Supplied coordinate may have missed this object, judging from
images in Fig. 19, top panel.}
\tablenotetext{h}{Large diameter is for the ring seen in X-rays; see Fig. 20, bottom panel and Chu et al. (2000).}
\tablenotetext{i}{This SNR is marginally detected only in C~III. This is the 
only detection listed with C~III but no O~VI emission.}
\end{deluxetable}  

%}
%\end{document}